# The Contribution of Low Surface-Brightness Galaxies to Faint Galaxy Counts

*Henry C. Ferguson*[1]
Space Telescope Science Institute [2]
3700 San Martin Drive, Baltimore, MD 21218
ferguson@stsci.edu

*Stacy S. McGaugh*
Institute of Astronomy, University of Cambridge
The Observatories, Madingley Road, Cambridge, CB3 0HA England
ssm@mail.ast.cam.ac.uk

# ABSTRACT

Recent observations have revealed a population of blue galaxies at intermediate redshift with a space density well in excess of expectations from the local luminosity function and standard cosmology. The colors and luminosities of these faint blue galaxies are similar to nearby low surface-brightness (LSB) galaxies. Such galaxies are severely underrepresented in surveys used to define the local luminosity function, but could *in principle* be detected in deep surveys. If LSB galaxy density is high enough, the faint-galaxy counts could be explained without requiring rapid galaxy evolution.

To explore the consequences of including LSB galaxies, we construct catalogs of simulated *non-evolving* galaxies drawn from a multivariate distribution of galaxy luminosities, central surface brightnesses, bulge/disk ratios and spectral-energy distributions. We compare two models dominated by LSB galaxies to a "standard" non-evolving model. Model galaxies are convolved with seeing and selected in a manner that closely matches real surveys. For each model we compute the local $B_J$ band luminosity function, HI mass function, number counts in the $B_J$, $I$, and $K$ bands, redshift distributions, and color distributions.

We find it possible to include a large population of LSB galaxies and incorporate a steep faint-end slope of the luminosity function in our simulations without violating the constraints on the local field-galaxy luminosity function or the HI mass function. For $q_0 = 0.5$, the most favorable model matches the counts to $B = 23$, but falls short of the observations at fainter magnitudes. The discrepancy at faint magnitudes is smaller in the $I$ and $K$ bands. The colors and redshift distributions remain roughly consistent with observations to $B = 24$. The most serious discrepancy with observations is in the distribution of $r_e$ at faint magnitudes, suggesting that the model contains too many LSB galaxies.

Nevertheless, the results suggest that LSB galaxies could be a significant contributor to faint-galaxy counts, reducing the need for such extreme models of galaxy evolution as rapid merging or bursting dwarf galaxies. We propose several tests to assess the contribution of LSB galaxies to faint galaxy counts and to differentiate models involving moderate galaxy evolution from models involving rapid merging or starbursts.

## 1. Introduction

Counts of very faint galaxies offer a simultaneous probe of the curvature of the universe and the evolution of its contents. The classic number–magnitude relation set out by Sandage (1961) as a test of the cosmological model has now been measured to the limits of 4 meter telescopes and the results are difficult to interpret. The counts of faint galaxies in the $B$ and $I$ bands show numbers are well in excess of cosmological models that do not include galaxy evolution, for any value of the deceleration parameter $q_0$ (Tyson & Jarvis 1979; Tyson 1988; Lilly, Cowie, & Gardner 1991; Metcalfe et al. 1991). Early attempts to reconcile the observations to the standard Friedmann-Lemaître cosmological model postulated that the faint-galaxy excess was primarily due to luminosity evolution: galaxies were brighter in the past because they were forming more stars (Tinsley 1980; Bruzual & Kron 1980; Yoshii & Takahara 1988; Guiderdoni & Rocca-Volmerange 1990). A robust prediction of such models is that the redshift distribution of faint galaxies should peak at higher $z$ than in non-evolving models. However, deep redshift surveys show a distribution that appears consistent with the no-evolution prediction to $B = 24$ (Broadhurst, Ellis, & Shanks 1988; Colless et al. 1990; Cowie, Songaila, & Hu 1991; Colless et al. 1993). Furthermore, while the $B$ counts show a strong excess over non-evolving models, counts in the $K$ band show no excess (Gardner, Cowie, & Wainscoat 1993).

Explanations offered for this discrepancy involve altering cosmology, modifying the standard picture of galaxy evolution, or altering properties of the local galaxy population. Fukugita et al. (1990) attempt to fit the count and redshift data with standard (Yoshii & Arimoto 1987) galaxy evolution models, and find that models with a sizable ($\lambda_0 \geq 0.5$) cosmological constant are favored. However, gravitational lensing statistics and Ly$\alpha$-cloud statistics (Fukugita & Turner 1991; Maoz & Rix 1993; Fukugita & Lahav 1991) favor a small or vanishing cosmological constant. Modifications of galaxy evolution include either rapid density evolution through merging or selective luminosity evolution. To explain the excess in the $B$ band, merger models require triggered bursts of star formation while the galaxies are still widely separated (Guiderdoni & Rocca-Volmerange 1991; Broadhurst, Ellis, & Glazebrook 1992; Lacey & Silk 1991; Lacey et al. 1993). The required merger rate is difficult to reconcile with constraints on the fraction of stars that could have formed in elliptical galaxies in the last 3 Gyr, the thinness of spiral galaxy disks, or the weak angular correlation among galaxies at $B = 25$ (Tóth & Ostriker 1992; Dalcanton 1993; Efstathiou et al. 1991; Roche et al. 1993). Models involving selective luminosity evolution propose that dwarf galaxies have faded more than giants in the last 3 Gyr (Broadhurst, Ellis, & Shanks 1988; Cole, Treyer, & Silk 1992; Lilly 1993). The most extreme models suppose that dwarf galaxy evolution is halted until redshifts $z < 1$, whereupon dwarfs form their stars in rapid bursts and subsequently fade beyond detectability (Babul & Rees 1992). For such a high space density of dwarf galaxies to have been missed locally requires that they have extremely low surface brightnesses, either because of expansion after expelling their gas in supernovae (Dekel & Silk 1986), or because of a top-heavy initial mass function.

A more conservative suggestion is that plausible modifications to the local luminosity function and distribution of galaxy spectral types can bring no-evolution (NE) models much closer to the data. Driver et al. (1994) present the limiting case of a dwarf-dominated no-evolution model, while Koo, Gronwall, & Bruzual (1993) primarily modify galaxy spectral-energy distributions. Neither model fully succeeds. Driver et al. (1994) attempt to match the counts by adopting a luminosity function with a very steep faint-end slope $\alpha = -1.8$. The resulting model is able to reproduce the faint-galaxy count and color data but predicts a median redshift well below that observed at $B = 22$ (0.18 for the model compared to 0.3 from the observations of Colless et al. 1993). Incompleteness in the combined LDSS survey (Colless et al. 1990; Colless et al. 1993) cannot be the cause of the discrepancy, as the survey is more than 95% complete. Koo et al. (1993) adopt a set of eleven plausible galaxy spectral types from the evolutionary models of Bruzual & Charlot (1993), then adjust the luminosity functions and space densities of the different types to provide the best simultaneous fit to the faint-galaxy counts, colors and redshift distributions. Post-facto comparison to the local field-galaxy luminosity function shows reasonable agreement to the limits of the Loveday et al. (1992) survey. The success of the model is due to the assumption of an open universe ($q_0 = 0.05$) and the inclusion of a much larger proportion of blue galaxies than typically seen in low-redshift surveys (see Fig. 8). Indeed, the model contains no galaxies redder than $B - V = 0.85$ with absolute magnitudes $M_{B_J} < -21$. This is equivalent to removing all ellipticals and S0 galaxies brighter than $L^*$ from the standard NE model.

Of course, within the standard cosmological framework, *all* galaxies must evolve. NE models are clearly unphysical and are intended only to provide a baseline to isolate the effects of the cosmological curvature (or selection effects) from the effects of galaxy evolution. Philosophically, we find the Koo et al. approach, which gives large weight to observations high-$z$ galaxies, a bit less useful than the standard approach of fixing the distribution of galaxy properties to match observations of nearby galaxies. In the standard NE model, it is easy to "turn on" evolution, adjusting the redshifts of formation, star-formation timescales, dust content, *etc.* to try to match the faint-galaxy observations for an assumed $q_0$ and $\lambda_0$, while still reproducing the properties of nearby galaxies. If this does not work (as many argue), then more exotic solutions are required. In the Koo et al. approach, it is not clear what form the evolution should take. There is no physically acceptable way to add "mild" evolution to their model. If one simply "turns on" evolution of galaxies at some high-$z$ with the star-formation timescales given in their Table 1, the luminosity function and color distribution will be a strong function of $z$, and the resulting distribution of nearby galaxy properties is unlikely to be acceptable. If one instead adopts their luminosity function and color distribution at $z = 0$ and evolves the models backwards with the assumed star-formation timescales, the predictions for faint galaxy counts and redshift distributions will change drastically; the three bluest classes of galaxies would form at redshifts $z < 0.5$, and hence disappear at the faint magnitudes where $k$-corrections would otherwise make them the most numerous population.

The model presented here is more in the spirit of standard NE models, with the exception that we propose a specific selection effect that could account for the discrepancy between the local galaxy properties and the deep counts. The crux of our model is the observation that the rest-frame isophotal limits of deep CCD surveys are fainter than the isophotal limits of the photographic surveys used to define the local luminosity function (McGaugh 1994). This is illustrated in Fig. 1, where we show the local and deep survey limits for three galaxies with the same total magnitude but with different scale lengths. Hence a large population of low-surface-brightness galaxies could contribute to the counts at faint magnitudes without violating current limits on the local luminosity function. While their local space density is not well known, examples of nearby LSB galaxies have been found in deep photographic surveys (Bothun et al. 1987; Davies, Phillipps, & Disney 1988; Schombert et al. 1990; Schombert et al. 1992). These galaxies are typically very blue (McGaugh & Bothun 1994), and appear to be weakly clustered (Bothun et al. 1993; Mo, McGaugh, & Bothun 1994), properties reminiscent of faint blue galaxies. LSB galaxies found in the field are typically HI-rich disk galaxies with low star-formation rates (McGaugh & Bothun 1994). The inferred color and luminosity evolution of such galaxies is slow enough that a non-evolving model is a reasonable approximation for their appearance at moderate redshifts.

The presentation is as follows. In §2 we briefly review the canonical wisdom on disk-galaxy surface brightnesses, and show that existing constraints do not rule out the possibility of a large population of LSB galaxies. Because the intrinsic distribution of surface brightness of disks is not well constrained, in subsequent sections we adopt three different distributions and compare their predictions for counts, redshift distributions, colors, and the HI mass function. The modeling is done via a Monte-Carlo technique described in §3. Comparisons to the observations are presented in §4. In §5, we discuss tests that can be carried out with high-resolution data to assess whether LSB galaxies are indeed a significant contributor to faint galaxy counts.

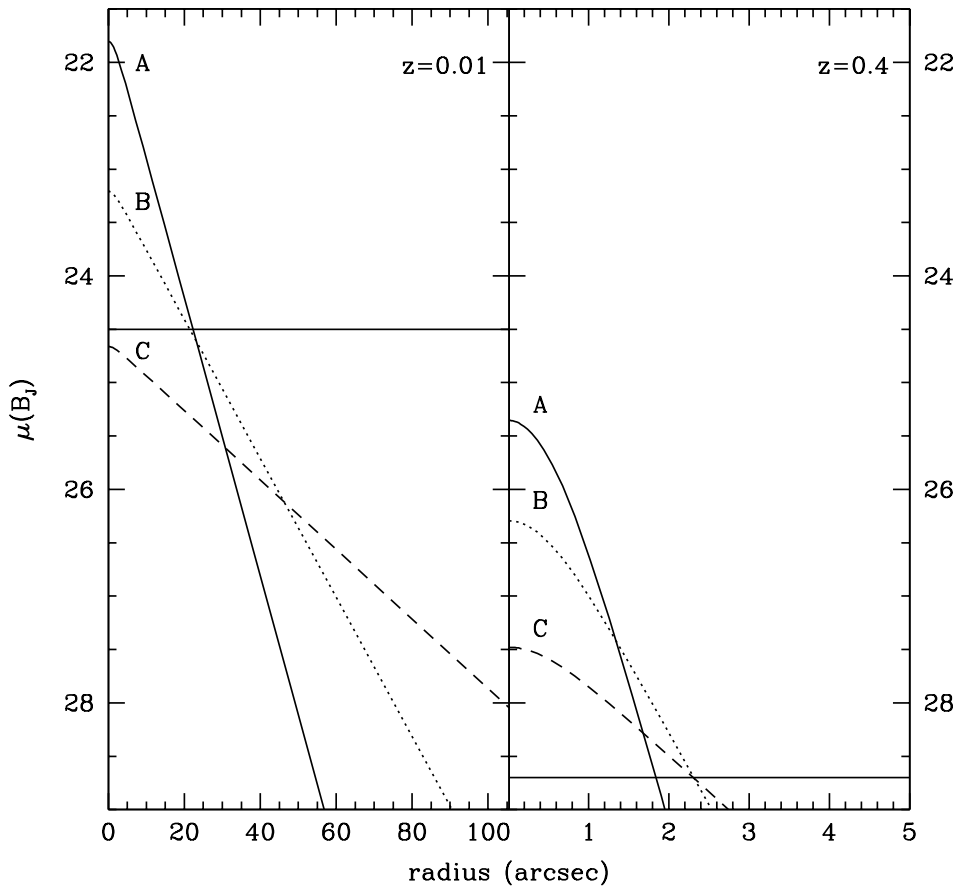

Figure 1:
This figure shows simulated radial surface-brightness profiles for three exponential disk galaxies with ($M_{B_J} = -19$). Galaxy A has the canonical Freeman central surface brightness $\mu_0(B_J) = 21.6$ at $z = 0$. Galaxies B and C have scale lengths a factor of 2 and 4 larger, respectively. The left panel shows the galaxies viewed in $3''$ (FWHM) seeing at $z = 0.01$. The horizontal line shows the isophotal threshold of APM survey (Loveday et al. 1992). Galaxy C would be missed entirely by the survey, while galaxy B would have a measured "total" magnitude too faint by 0.7 mag, if the isophotal-to-total correction were based on galaxy A. The right panel shows the same three galaxies shifted to $z = 0.4$ and viewed in $1''$ seeing. A type II (see text) SED was assumed, giving 1.07 mag of $k$ dimming. The horizontal line in this panel is Tyson's (1988) isophotal threshold $\mu_{B_J} = 28.7$. All three galaxies would be detected above this threshold. Isophotal-to-total magnitude corrections are 0.1, 0.2, and 1.0 mag for galaxies A, B, and C, respectively.

## 2. The Intrinsic Distribution of Central Surface Brightness of Galaxy Disks

Most determinations of the galaxy luminosity function and number counts predicted therefrom implicitly assume that the distribution of galaxy surface brightnesses is a $\delta$-function. This is an important simplification, as it allows one to make use of easily measured isophotal magnitudes without the need for detailed surface photometry, and suppresses one dimension of integration. It is observationally supported by Freeman (1970), who found that all spiral galaxies had central surface brightnesses $\mu_0 = 21.65 \pm 0.35 B$ mag arcsec$^{-2}$, the scatter being consistent with observational error.

Disney (1976) pointed out that selection effects could cause the *apparent* distribution to appear sharp even if the intrinsic distribution were broad. Allen & Shu (1979) concurred that such selection effects could act against surface brightnesses fainter than the Freeman value, but argued that higher surface brightness objects would not be missed. Disney & Phillipps (1983) developed a formalism to correct the apparent distribution in which a particular value of the central surface brightness is favored, and both high and low surface brightness objects can be missed.

That the apparent distribution of central surface brightnesses peaks at the Freeman value has been confirmed by Phillipps et al. (1987) and van der Kruit (1987) who found $\mu_0 = 21.75$ and $21.5$ (in $B_J$), respectively. These authors employed different methods to recover the intrinsic distribution. Phillipps et al. (1987) applied the method of Disney & Phillipps (1983), and found a distribution which was broad and asymmetric. Davies (1990) modeled the way in which central surface brightnesses were measured by Phillipps et al. (1987) and concluded that deviations from pure exponential profiles caused by even modest bulge components would broaden the distribution still further, implying large numbers of LSB galaxies. The V/V$_{max}$ method (Schmidt 1968) was used by van der Kruit (1987), who found that a narrow scatter of 0.4 mag. about the Freeman value was recovered if the sample was restricted to large, early type galaxies. However, the distribution did not approximate a $\delta$-function for dwarf galaxies.

The results of these studies are mutually inconsistent, and the situation remains confused. What is really needed is the bivariate distribution of luminosity and surface brightness. Since the surface brightness portion of the distribution is not well constrained, we adopt two very different forms of the bivariate distribution in order to demonstrate the effects on the predicted counts. The first (model A) increases the normalization of the luminosity function, while the second (model B) steepens the slope of the faint end. This is accomplished with-out violating constraints on the local luminosity function when the procedures of isophotal measurement and selection are taken into account.

In model A, galaxies exist with equal numbers per unit magnitude over the range of central surface brightness $21.6 \leq \mu_0 \leq 25$ in $B_J$. This is motivated by the existence of large, low surface brightness galaxies (Bothun et al. 1987; Schombert & Bothun 1988; Impey & Bothun 1989; Schombert et al. 1992). While the space density of such galaxies is not well known, it no doubt exceeds that expected from a Gaussian distrubution of central surface brightness with a dispersion $\sigma = 0.4$ mag about the Freeman value. In order to isolate the effect of this surface brightness distribution on one parameter of the luminosity function (the normalization) it is assumed that there is no correlation between surface brightness and luminosity (i. e., the shape of the luminosity function is the same for every surface brightness). This is consistent with, though not demanded by, the similarity between the HI rotation velocity distributions of LSB and other field galaxies, which suggests that they all have comparable masses (Schombert et al. 1992).

The range of surface brightnesses included in the flat distribution of the model is dictated at the faint end by the fact that galaxies fainter than $\mu_0 \sim 25$ will not contribute significantly to the counts even if large and luminous. At the bright end the distribution is truncated at the Freeman value following the results of Allen & Shu (1979) that very few high surface brightness galaxies exist. *If* there are many of these, then of course cosmological dimming will bias faint galaxy samples towards high surface brightnesses (Phillipps, Davies, & Disney 1990).

In model B, we assume that the narrow distribution of central surface brightnesses found by van der Kruit (1987) effectively holds for giant $(L > L^*)$ galaxies, but that there is a systematic trend between luminosity and surface brightness for fainter galaxies. This strongly affects the slope of the faint end of the luminosity function when isophotal rather than total measures are used because intrinsically faint galaxies have their luminosities significantly underestimated. To illustrate the severity of this affect, we assume the following:

$$\mu_0 = 21.6 \pm 0.4 \quad \text{for} \quad L \geq L^* \tag{1}$$

and

$$\mu_0(L) = 21.6 - 2.5\log(L/L^*) \pm 0.4 \quad \text{for} \quad L < L^*. \tag{2}$$

The equation for sub-$L^*$ galaxies is essentially a constant size relation, with some (Gaussian) scatter.

Finally, for comparison, we compute a "standard" NE model using a galaxy mix similar to that adopted by Yoshii & Takahara (1988), but incorporating isophotal selection with disk central surface brightnesses set to the Freeman value with 0.4 mag Gaussian scatter. These distributions are illustrated schematically in Fig. 2.

Models A and B are clearly *ad hoc*, and are intended not to represent reality but to illustrate the importance of galaxy selection criteria and magnitude estimation techniques in both the local and deep surveys. Model B is perhaps closer to the truth in that it includes the general trend of decreasing surface brightness with decreasing luminosity observed for Virgo and Fornax Cluster dwarf galaxies (Binggeli, Sandage, & Tarenghi 1984; Ferguson

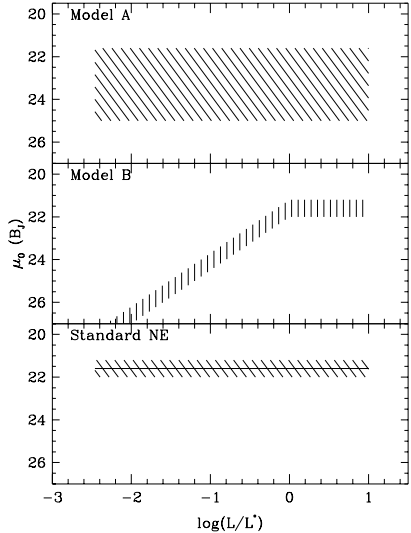

Figure 2: Schematic illustration of the central-surface-brightness distributions assumed for galaxy disks in our models. In model A, we assume that surface-brightness is independent of luminosity. In model B, surface brightness depends on luminosity as described in the text, with 0.4 mag scatter about the mean relation. The Monte-Carlo NE model assumes Freeman disks with 0.4 mag scatter. The analytic NE model assumes no scatter.

& Sandage 1988). The surface-brightness–luminosity relation in model B is much steeper than observed, but is difficult to rule out as galaxies with low surface brightness for their luminosity will always be preferentially missed in real surveys. An intermediate case might be one with a shallower trend of surface brightness with luminosity but with broader scatter in central surface brightness at fixed luminosity (as found by Impey, Bothun, & Malin 1988 and Bothun, Impey, & Malin 1991). Model B is also similar to the model of Lacey et al. (1993), which included isophotal selection with a surface-brightness–luminosity relation similar to that seen in Virgo and a steep luminosity function. These properties were *predicted* from their model of tidally triggered galaxy formation that included halo formation, galactic winds, spectral evolution, and extinction. The effects of surface brightness are thus difficult to disentangle from other aspects of the model.

## 3. Construction of Monte-Carlo Models

The simplest way to include intrinsic scatter into the galaxy distribution functions is to construct simulated galaxies using Monte-Carlo techniques. For the three models described above galaxy parameters are chosen at random from the distribution functions describing space density (constant), luminosity, surface brightness, morphological type, and bulge/disk ratio. Catalogs of simulated galaxies are constructed and are observed (i.e. they are selected and their magnitudes are measured) in a way that closely matches real surveys.

For comparison, and as a check on our Monte-Carlo technique, we compute a second

Table 1: Galaxy Parameters

NE model

| Type | $M_B^*$ | $\alpha$ | $\phi^*$ ($10^{-3}$Mpc$^{-3}$) | bulge to total | Bulge SED | Disk SED |
|---|---|---|---|---|---|---|
| E/S0 | -21.1 | -1.1 | 0.32 | 1.0 | I | |
| S0 | -21.1 | -1.1 | 0.58 | 0.4 ± 0.10 | I | I |
| Sab | -21.1 | -1.1 | 0.58 | 0.3 ± 0.08 | I | II |
| Sbc | -21.1 | -1.1 | 0.58 | 0.15 ± 0.04 | I | III |
| Sdm | -21.1 | -1.1 | 0.40 | 0.0 | | III |

Model A

| Type | $M_B^*$ | $\alpha$ | $\phi^*$ ($10^{-3}$Mpc$^{-3}$) | bulge to total | Bulge SED | Disk SED |
|---|---|---|---|---|---|---|
| E/S0 | -21.1 | -1.0 | 0.35 | 1.0 | I | |
| S0 | -21.1 | -1.1 | 0.71 | 0.4 ± 0.10 | I | I |
| Sab | -21.1 | -0.5 | 0.97 | 0.3 ± 0.08 | I | II |
| Sbc | -21.1 | -0.5 | 0.97 | 0.15 ± 0.04 | I | III |
| Sdm | -20.5 | -1.2 | 0.13 | 0.0 | | III |

Model B

| Type | $M_B^*$ | $\alpha$ | $\phi^*$ ($10^{-3}$Mpc$^{-3}$) | bulge to total | Bulge SED | Disk SED |
|---|---|---|---|---|---|---|
| E/S0 | -21.1 | -0.5 | 0.44 | 1.0 | I | |
| S0 | -21.0 | -0.5 | 0.91 | 0.4 ± 0.10 | I | I |
| Sab | -20.9 | -0.5 | 1.21 | 0.3 ± 0.08 | I | II |
| Sbc | -20.7 | -1.0 | 1.48 | 0.15 ± 0.04 | I | III |
| Sdm | -20.6 | -1.8 | 1.44 | 0.0 | | III |

NE model with total magnitude selection, using the analytic form of the luminosity function and numerically integrating over the luminosity function to compute $N(m)$ and $N(z)$ distributions. The techniques are identical to those used by Yoshii & Takahara (1988), but the luminosity function and type distributions are slightly different. For our comparison we use luminosity functions identical to those for the isophotal NE model. The distribution of spectral-energy distributions (SED's) is slightly different, however, as the Yoshii & Takahara models do not include separate bulge and disk components.

Galaxies are divided into five broad Hubble types, each characterized by the following:

1. a Schechter (1976) luminosity function

$$\phi(L)dL = \phi^*(L/L^*)^\alpha e^{-L/L^*} d(L/L^*), \qquad (3)$$

characterized by a space density $\phi^*$, a faint-end slope $\alpha$, and a characteristic luminosity $L^*$;

2. a ratio of bulge/total luminosity in the $B$ band (Simien & de Vaucouleurs 1986); and Gaussian scatter about this ratio characterized by a dispersion $\sigma$ (different for each type) and constrained such that $0 \leq L_B(bulge)/L_B(total) \leq 1$;

3. surface brightness profiles $g(r)$ given by

$$g(r) = g_0 \exp(a_n(r/r_e)^{1/n}), \qquad (4)$$

with $n = 1$ for galaxy disks and $n = 4$ for galaxy bulges and coefficients $a_1 = 1.68$ and $a_4 = 7.67$; and

4. separate (luminosity independent) spectral-energy distributions (Coleman, Wu, & Weedman 1980) for the bulge and disk components.

The parameters describing the different models are shown in Table 1.

For the Monte-Carlo models, we construct simulated catalogs of galaxies for each morphological type, selecting the parameters for each galaxy at random from the distribution functions of bulge/total ratio, central surface-brightness, and luminosity. The code uses a double-precision random number generator, and has been extensively tested to ensure that input distribution functions are properly reproduced by the random selection, even in the tails of the distribution. Redshifts of the galaxies are selected so as to produce a uniform co-moving density using the rejection method (Press et al. 1992) with the Euclidean volume element as the comparison function to the cosmological volume element for $q_0 = 0.5$. Magnitudes are then computed for specified bandpasses by integrating the redshifted spectral-energy distributions through the filter bandpasses. Depending on the survey we are trying to simulate, these catalogs contain anywhere from $10^3$ to $10^5$ galaxies, and list for each galaxy the redshift, luminosity in the rest-frame $B_J$ band, apparent magnitudes in various bands, and scale lengths of the bulge and disk components. These galaxy catalogs are then fed to separate programs that "observe" the galaxies using seeing and selection criteria that closely match real surveys. Distributions of apparent magnitude, redshift, color, etc. are then compiled from galaxies that pass the selection criteria, using isophotal, aperture, or total magnitudes as appropriate. The resulting distributions are normalized to represent the correct volume densities of each galaxy type. Note that our simulations do not explicitly include the noise present in any real observations. To the extent that the algorthims used in the deep surveys for detecting galaxies and measuring their magnitudes are unbiased, the effect of noise is simply to increase the scatter in the measured magnitudes. Scatter of a few tenths of a magnitude is unimportant over the many decades of the $N(m)$ diagram and over the relatively wide magnitude intervals used for $N(z)$. In any case, the best way to assess the impact of noise would be to construct simulated images with noise and analyze them in the exact same way as the observations. While such an experiment would be worthwhile, it is beyond the scope of this paper.

Table 2: Assumed Survey Parameters

| Band | Observer | Magnitude | Seeing ($''$) | $D_{min}$ ($''$) | $\mu_{\lim}$ | |
|---|---|---|---|---|---|---|
| $B_J$ | Loveday | Isophotal | 3.0 | — | 24.5 | 1 |
| $B_J$ | KOS78 | Total | 3.0 | 20 | 24.5 | 2 |
| $B_J$ | Tyson | Isophotal | 1.7 | — | 28.7 | |
| $B_J$ | Colless | Isophotal | 1.0 | — | 26.5 | 3 |
| $B_{AB}$ | Cowie | Isophotal | 1.0 | 1.0 | 28.6 | 4 |
| $I_{AB}$ | Lilly N(m) | Hybrid | 1.2 | 2.0 | 28.0 | 5 |
| $I_{AB}$ | Lilly N(z) | Aperture | 0.7 | 3.0 | 28.0 | 6 |
| $K_{AB}$ | Cowie | Aperture | 1.0 | 3.0 | 25.4 | 7 |

Notes:
1. Correction -0.27 added to isophotal magnitudes.
2. Rough estimates. Used to set type fractions.
3. Correction -0.58 added to isophotal magnitudes.
4. Assumed complete to $D_{28.6} > 2''$. Isophotal magnitudes used.
5. Isophotal magnitudes for galaxies with $D_{28} > 2''$; otherwise, used $2''$ aperture.
6. Assumed complete to $D_{28} > 2''$. $3''$ aperture magnitudes used.
7. Assumed complete to $D_{25.4} > 1''$. Isophotal magnitudes used.

The surveys we have chosen to simulate are listed in Table 2. Each is described by the estimated FWHM of the seeing disk, a type of selection (diameter or magnitude), and a type of magnitude (aperture, isophotal, hybrid, or total). We *tune* the model to match roughly the canonical distribution of morphologies and total luminosity function in bright surveys (Table 2 lines 1 & 2). This involves adjusting $\phi^*$, $L^*$, and $\alpha$ separately for the different Hubble types. As an independent test of the properties of local galaxies, in §4.2 we predict the color distributions for a sample of bright galaxies and compare to the observed distribution for a subsample of the RC3 catalog (de Vaucouleurs et al. 1991). We then compute the faint galaxy counts, redshifts, and colors for the rest of the surveys listed in Table 2.

### 3.1. *Cosmological Model*

Our models are constructed in the framework of the standard Friedman-Lemaître model with zero cosmological constant. The relevant formulae are set out in Yoshii & Takahara (1988) and Sandage (1988). Briefly, the apparent magnitudes of sources in a bandpass $\lambda$ are related to their absolute magnitudes $M_\lambda$ by

$$m_\lambda = M_\lambda + k_\lambda(z) + 5\log(d_L) + 25, \qquad (5)$$

where $k_\lambda$ is the standard $k$-correction that incorporates the frequency shift and the bandpass dilation due to redshift and $d_L$ is the luminosity distance in Mpc. The luminosity distance is

$$d_L = \frac{c}{H_0 q_0^2}\{q_0 z + (q_0 - 1)[\sqrt{1 + 2q_0 z} - 1]\}, \qquad (6)$$

where $H_0$, $q_0$, and $c$ are the Hubble constant, the deceleration parameter, and the velocity of light, respectively. As our aim in this paper is to elucidate the effects of surface-brightness selection, rather than to test a particular cosmology, we adopt $H_0 = 50$ km s$^{-1}$Mpc$^{-1}$ and $q_0 = 0.5$ throughout.

The co-moving density of galaxies is conserved in our model. The number of galaxies per unit redshift per steradian depends only on the volume element

$$\frac{dV}{dz} = \frac{cd_L^2}{H_0(1+z)^3\sqrt{1+2q_0z}}. \tag{7}$$

Galaxy angular sizes are computed using the angular-diameter distance

$$d_A = \frac{d_L}{(1+z)^2}. \tag{8}$$

### 3.2. Point-Spread Function Convolution

Atmospheric distortions (seeing) can have a significant effect on galaxy counts at faint magnitudes. This effect was not included in McGaugh's (1994) initial consideration of the counts of LSB galaxies, but has been considered in other contexts by Pritchet & Kline (1981) and Yoshii (1993). To compare our results to the counts from deep surveys, we convolve our model galaxy profiles with point-spread functions that closely match the conditions in the real surveys. Assuming a circular galaxy with a surface brightness profile $g(\beta)$, where $\beta = r/r_e$, and assuming a circular Gaussian point-spread function with dispersion $\sigma$ in units of $r_e$, the convolved one-dimensional profile has the form

$$\tilde{g}(\sigma,\beta) = \sigma^{-2}f(\beta)\int_0^\infty g(\xi)I_0(\beta\xi/\sigma^2)f(\xi)\xi d\xi. \tag{9}$$

$I_0(x)$ is the modified Bessel function of the first kind and

$$f(x) = \exp(-x^2/2\sigma^2). \tag{10}$$

To speed our calculations we have computed the surface-brightness profile $\tilde{g}(\sigma,\beta)$ and the integrated profile

$$\tilde{G}(\sigma,\beta) = 2\pi \int_0^\beta \tilde{g}(\sigma,\xi)\xi d\xi \tag{11}$$

separately for bulge and disk profiles for a grid of $0 \leq \beta \leq 10$ and $0 \leq \sigma \leq 10$ in steps of 0.1 in $\beta$ and $\sigma$. For each model galaxy we compute $r_e$ in arcsec for the disk and bulge components using the angular-diameter distance, and compute $g(0)$ through the filter bandpass using the appropriate $k$ corrections. To compute an isophotal radius, we step through the grid $\tilde{g}$, using a value of $\sigma$ that approximates real observing conditions, and summing bulge and disk components until the flux drops below that corresponding to the limiting isophote.

### 3.3. *The Distribution of Morphological Types*

When isophotal selection is not considered, the actual morphologies of galaxies are irrelevant and all that is important is their spectral energy distributions. A typical assumed mix of Hubble types is 30% E/S0, 50% Sa-Sb, and 20% Sc or later (Tinsley 1980; Shanks et al. 1984; Yoshii & Takahara 1988). This morphological mix is justified on the basis of wide-area surveys of bright galaxies such as those of Tammann, Yahil, & Sandage (1979); Kirshner, Oemler, & Schechter (1978); and (Peterson et al. 1986).

As our models include isophotal selection, both spectral energy distributions and galaxy profiles are important. Galaxies in our models are simple entities composed of pure $r^{1/4}$-law bulges and pure exponential disks. We have separated galaxies into five morphological classes. For each type, we fix the ratio of bulge to total light in the $B_J$ band to mean values found by Simien & de Vaucouleurs (1986), but allow Gaussian scatter about this mean. The adopted bulge/total parameters are shown in Table 1. Spectral energy distributions are specified separately for bulge and disk components (see below), so the ratio of bulge/total light will vary with bandpass.

The key new feature of our model is the inclusion of LSB disk galaxies. Matching the distribution of morphological types in wide-area surveys introduces an additional uncertainty in our models, as the selection criteria of those surveys are not well quantified. In particular the fraction of late-type galaxies (especially in model B) is critically dependent on the limiting isophote of the surveys. As these limits have never been explicitly quantified for surveys with good morphological resolution, we have simply chosen a limit that appears to us to be a good approximation of the survey material used by Kirshner et al. (1978) and Shanks et al. (1984). Specifically, we use our code to simulate a survey in $3''$ seeing complete to $B_J = 16$ for galaxies with isophotal diameters $D_{24.5} > 20''$. We adjust the space densities of the different types to produce the proportions E:S0:Sab:Sbc:Sdm = 10:20:25:25:20 in the final survey.

### 3.4. *The Luminosity Function*

While each morphological type is characterized by its own luminosity function, for the overall normalization we require that the total luminosity function match that observed locally (Kirshner, Oemler, & Schechter 1979; Efstathiou, Ellis, & Peterson 1988; Loveday et al. 1992). Our goal here is to demonstrate that when a realistic model for isophotal selection is included, existing observations still allow room for a large population of LSB galaxies. To model the local luminosity function (LF), we try to mimic the selection criteria of Loveday et al. (1992; herafter LPEM). However, for comparison to previous faint-galaxy modeling, we have adopted the Yoshii & Takahara (1988) LF slope ($\alpha = -1.1$) and normalization ($\phi^* = 2.3 \times 10^{-3} \, \mathrm{gal \, Mpc^{-3}}$) in preference to the best fit found by LPEM. Our adopted local LF is consistent to their best fit to within their $1\sigma$ errors; adopting the LPEM best fit would increase the discrepancy in $N(m_{B_J})$ for all models, but would not affect the comparison of the different models.

The LPEM field-galaxy luminosity function was derived from a redshift survey of galaxies detected on IIIaJ Schmidt plates by the Automatic Plate Measuring (APM) machine (Kibblewhite et al. 1984). An automated star–galaxy separation algorithm was used, combined with visual inspection of the images, to decide whether or not to include an object in the survey. Sources were detected by the APM above an isophote of $\mu_{B_J} \sim 24.5 \,\mathrm{mag\, arcsec^{-2}}$ (Loveday 1989). APM magnitudes are simply a logarithmic scaling of the sum of the linearized pixel intensities above this threshold. They therefore closely correspond to isophotal magnitudes, with the exception that saturation at surface brightnesses $\mu_{B_J} < 22$ may artificially increase the magnitudes of high-surface-brightness galaxies. Loveday (1989) used CCD images of selected galaxies to compute a constant conversion from APM magnitudes to total $B_J$ magnitudes. This process makes no correction for the different proportions of the total galaxy light encompassed for high and low surface-brightness galaxies above the APM detection threshold [3]. The magnitudes are therefore essentially isophotal, with a constant offset.

To simulate this process, we compute isophotal magnitudes at $\mu_{B_J} = 24.5$ for each of our simulated galaxies (assuming 3″ FWHM seeing), but then add a constant offset of $\Delta m = -0.27$ mag to simulate the conversion from APM to "total" magnitudes. This is the conversion appropriate for pure Freeman disks, which presumably dominate the calibration sample. We select galaxies from our simulated catalogs with these "corrected" magnitudes in the range $15 < B_J < 17.15$, and use the standard $V/V_{max}$ technique (Schmidt 1968) to reconstruct the luminosity function. We tune the shapes of the type-specific luminosity functions to produce an "observed" luminosity function that matches a Schechter function with $\phi^* = 2.3 \times 10^{-3} \,\mathrm{gal\, Mpc^{-3}}$ and $\alpha = -1.1$. Figures 3-5 show the comparison of the models to our adopted field-galaxy LF. The rather steep intrinsic faint end slope of the model B luminosity function is consistent with that determined by Bothun, Impey, & Malin (1991) from the distribution of scale lengths and surface brightnesses found in the Fornax Cluster (not shown in the figure).

### 3.5. Spectral Energy Distributions

We use standard spectral-energy distributions (SED's) determined from observations of nearby galaxies as the basis for our magnitude computations (Pence 1976; Coleman, Wu, & Weedman 1980). While these SED's undoubtedly do not cover the full range exhibited by real galaxies, we take them as a conservative starting point. In this respect, our model is distinctly different from that of Koo et al. (1993), where much of the increase in counts comes from galaxies that are bluer than commonly found in local galaxy samples. Our ability to match the counts without producing excess blue galaxies locally illustrates that surface-brightness selection effects are at least as important as $k$-corrections in governing the counts.

---

[3] Though Loveday (1989) searched for the possibility a surface brightness effect, the bright isophotal limit does not allow sufficient dynamic range for this effect to be readily apparent, especially when there is large scatter in the calibration data. Indeed, much of the scatter in the raw APM magnitudes could be due to fluctuations in the isophotal limit from plate to plate. In an ideal situation, correction from isophotal to total magnitudes should be made on an individual basis for each galaxy based on profile shape. When a single average correction is applied instead, the most seriously affected objects are those associated with large volume corrections, producing a pronounced effect on the luminosity function.

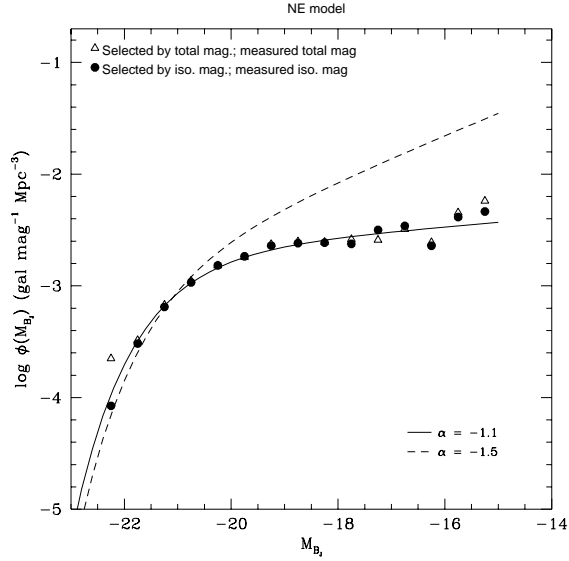

Figure 3: This and the next two figures show the luminosity functions for our models. The solid line shows our fiducial "observed" total luminosity function, consistent with the APM survey of Loveday et al. (1992). For comparison, the dashed line shows a luminosity function with the same $\phi^*$, but with $\alpha = -1.5$. The open symbols show the luminosity function that would be recovered from our simulated galaxy catalogs if galaxies could be selected by total magnitude. The solid symbols show the luminosity function recovered when the APM selection criteria and magnitude estimation scheme are adopted as described in the text. The luminosity functions of the individual morphological types have been tuned to match the fiducial total luminosity function, and the three panels here show the extent to which this tuning has been successful. This panel shows the NE model.

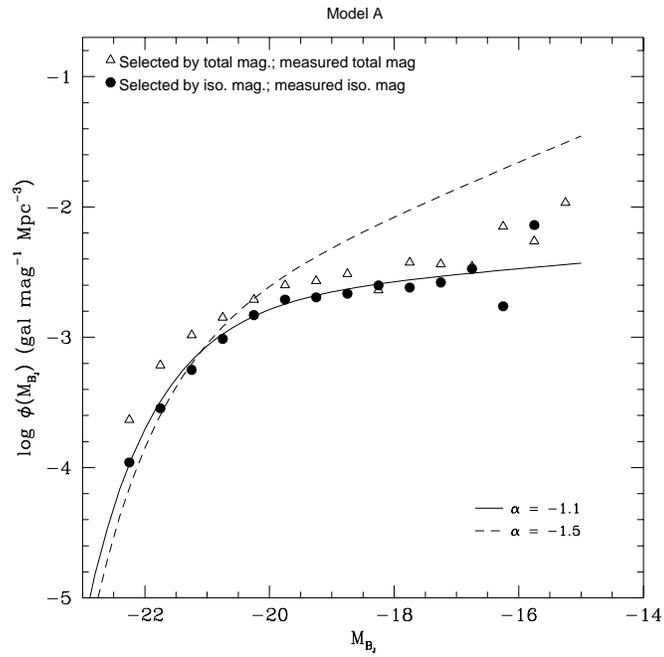

Figure 4: Model A.

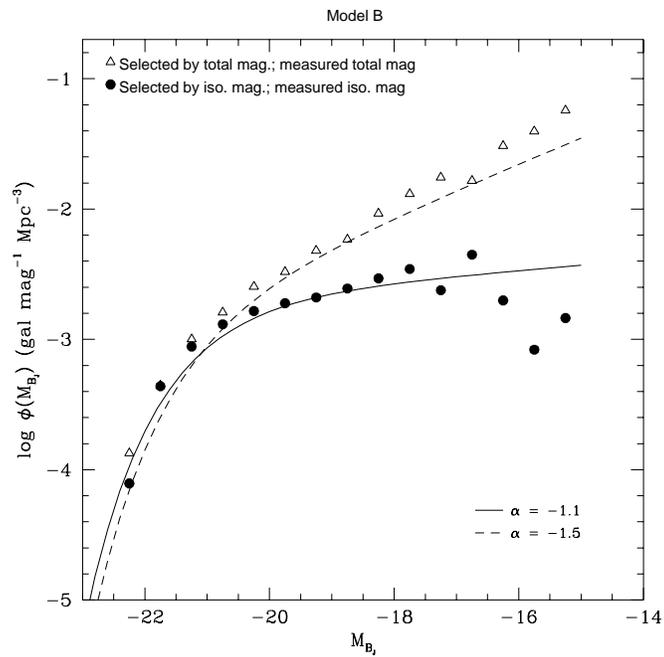

Figure 5: Model B.

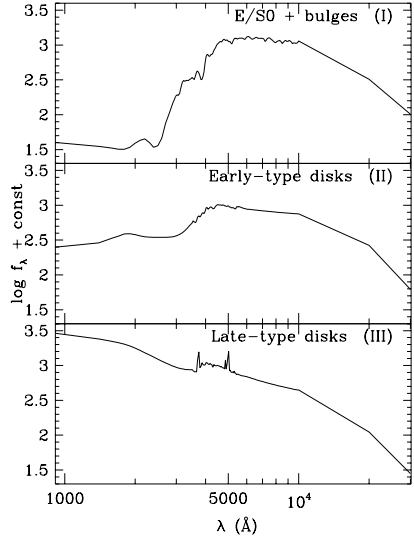

Figure 6: Adopted spectral energy distributions.

Table 3: Photometric Zero Points
$10^9 f_\lambda \, \mathrm{erg\, cm^{-2} s^{-1} \mathring{A}^{-1}}$ for $m_\lambda = 0$

| Band | Zeropoint |
|---|---|
| $B_J$ | 5.27 |
| $B$ | 6.29 |
| $V$ | 3.64 |
| $B_{AB}$ | 5.61 |
| $I_{AB}$ | 1.34 |
| $K_{AB}$ | 0.22 |

The SED's adopted are shown in Fig. 6. For $1400 \leq \lambda \leq 10000\,\mathring{A}$ we use the compilation of Coleman, Wu, & Weedman (1980). We adopt three of their SED's, labeling them I, II, and III. Type I is used for ellipticals, bulges and S0 disks, and is taken from their Table 2 tabulation of the mean M31 + M81 SED. Type II is is taken from their Table 3 and is used for Sab disks. Type III comes from their Table 5 SED for Im galaxies. These SED's are extrapolated to long wavelengths following Yoshii & Takahara (1988) and linearly extrapolated to short wavelengths as shown in Fig. 6. The SED's chosen for the bulge and disk components of each galaxy type are shown in Table 1. We normalize the flux in each SED to produce an absolute magnitude $M_{BJ} = -21.1$ in the rest-frame, the multiply by the appropriate factor for the absolute magnitude of the disk or bulge. We then redshift the galaxy and compute a weighted mean flux over the filter bandpass to account for $k$-dimming.

To convert to magnitudes we use the zero points shown in Table 3.

### 3.6. HI Mass Function

A standard argument against a high space-density of LSB galaxies is that blind HI surveys do not find many optically invisible HI clouds (Ferguson & Sandage 1988; Briggs 1990). We regard this as a serious constraint, and devote §4.1 to a detailed discussion of the HI constraints. To compare our luminosity function to HI surveys, we need a relation between optical luminosities and HI masses. We assume E and S0 galaxies have no HI, and adopt for later types the conversion formula of Briggs (1990):

$$M_{HI} = 3.2 \times 10^9 M_\odot (L/L_0)^{0.9}, \qquad (12)$$

where we adopt $L/L_0 = 10^{-0.4(M_{B_J}+21.1)}$ for all types regardless of the value of $L^*$ used for the optical luminosity function. Comparison to the observed HI mass function is shown in Fig. 7 and discussed in §4.1.

### 3.7. Galaxy Surface-Brightness–Luminosity Relations

The most important departure of our model from others that have considered isophotal selection is that we adopt non-standard models for the distribution of central surface brightness of spiral galaxy disks (for galaxies of type Sa and later). Our assumptions are described in §2 and illustrated in Fig. 1.

The disks of S0 galaxies are assumed to have $\mu_0 = 21.6 \pm 0.4$, regardless of luminosity. Because we are ignoring evolution, the properties of S0 disks are not important for our models, as the $k$ corrections preferentially remove early-type galaxies from high-redshift samples.

For elliptical galaxies and bulges, we use the relation from Sandage & Perelmuter (1990)

$$\mu_e = -0.48\, M_{B_T} + 11.02, \qquad (13)$$

which translates to

$$\mu_e = 1.20 \log(L/L*) + 21.16 \qquad (14)$$

for $M_{B_T}{}^* = -21$. In this model, surface brightness increases with decreasing luminosity for elliptical galaxies. While this appears to hold for $r^{1/4}$-law ellipticals seen in nearby clusters, it does not hold for the dwarf-ellipticals that dominate the counts in those clusters (Binggeli, Sandage, & Tarenghi 1984; Ferguson & Sandage 1988). However, dE galaxies do not appear to be abundant in the field, and are in any case bluer than giant E's and have exponential surface-brightness profiles. Thus, to the extent that dE galaxies are included, they are grouped implicitly with low-luminosity LSB galaxies in our models. The E and S0 galaxy luminosity functions in our models are either flat or declining at faint magnitudes, and hence do not predict large numbers of compact high-surface brightness galaxies at low luminosities. Isophotal selection makes very little difference to the counts of early-type galaxies in our models.

### 3.8. Galaxy Selection Functions in Deep Surveys

The details of object selection and photometry in faint galaxy surveys are difficult to model precisely and are sometimes not completely specified in the published reports. Automated galaxy detection algorithms typically catalog pixels above a certain S/N threshold, then assign adjacent detected pixels to a single "object." To increase the detection probability for faint galaxies, the images are usually convolved with a kernel of 1-2″ FWHM before running the detection algorithm. Various star-galaxy separation procedures are used to remove unresolved objects, but the details of these are unimportant at faint magnitudes, where stars are a negligible fraction of the total source counts. For the detected objects, some surveys report isophotal magnitudes, others aperture magnitudes, and still others report a hybrid of the two.

The ultimate test of the models would be to construct simulated images with signal-to-noise ratios and PSF's that match the real surveys, then analyze them with the same software that was used on the real images. This task is beyond the scope of our paper, and in any case is best done in collaboration with the actual observers. For illustrative purposes, we present in §5.4 simulated deep HST WFPC-2 images, but for comparison to existing observations we use the approximations to the surveys described below.

For the $B_J$-band counts, we compare to Tyson (1988). His isophotal threshold was $\mu_{B_J} = 28.7$ and he reports isophotal magnitudes above that threshold. The seeing FWHM was about 1.7″. We use the Lilly et al. (1991) survey for the $I$ band. Magnitudes used for the $N(m)$ diagram in that study were in the Oke (1974) AB system ($I_{AB} = I + 0.48$) and were isophotal above a threshold of $\mu_{I_{AB}} = 28$ for galaxies with $D_{28} > 2″$, but were aperture magnitudes through a 2″ aperture for smaller galaxies. Images were convolved to an effective seeing of 1.2″. The $K$-counts are taken from Gardner, Cowie, & Wainscoat (1993). Details of the galaxy selection and photometry for the deepest $K$ counts have not been published. We assume, following Yoshii (1993), that the survey is complete for galaxies with $K_{AB}$ ($= K + 1.8$) isophotal diameters $D_{25.4} > 1″$. The seeing FWHM was taken to be 1.7″.

At the faintest levels, the published counts include completeness corrections for the number of overlapping sources. We have not simulated this process, as our model galaxies are all counted if they meet the survey selection criteria.

We compare our redshift distributions to three published samples (Cowie, Songaila, & Hu 1991; Lilly 1993; Colless et al. 1990, 1993). Deep Anglo-Australian Telescope (AAT) prime-focus plates were used to select galaxies for the LDSS redshift survey (Colless et al. 1990, 1993). The magnitudes measured were essentially isophotal above $\mu_{B_J} \sim 26.5$, with a constant (unspecified) shift to bring them into agreement with a standard sequence of galaxies measured through a 10″ aperture with a CCD. An Sbc Freeman disk observed at $z = 0.4$ would have a central surface brightness of $\mu_0(B_J) = 24.1$ In 2″ seeing the correction from $m_{26.5}$ to $m(10″)$ would be 0.58 mag, which we adopt for all galaxies in the sample. The Lilly (1993) sample was compiled from deep CCD imaging with the Canada-France-Hawaii Telescope (CFHT) in 0.7″ seeing. The isophotal selection limits are not specified. For our models, we assume that the sample is complete for galaxies with $D_{28} > 2″$. Magnitudes for the detected galaxies are computed through a 3″ aperture, and galaxies are required to have aperture magnitudes $21.0 < I_{AB} < 22.5$. The deepest redshift survey is that of Cowie, Songaila, & Hu (1991), who published redshifts for a small sample of galaxies to $B = 24$ selected from the Lilly, Cowie, & Gardner (1991) survey. The selection criteria of the final sample of 21 objects are not well defined, as only 13 were from the original input sample. In our models, we assume the survey was diameter-limited at $D_{28.63}(B_{AB}) > 2''$, and use uncorrected isophotal magnitudes to decide whether the galaxies are brighter than the magnitude limit.

While the galaxy counts are very sensitive to selection criteria and magnitude schemes in our models, the redshift distributions are not. In the NE model and model A the luminosity function is not steeply rising, so a change in the limiting isophote does not bring in many low-redshift LSB galaxies. The steeply rising luminosity function in model B, however, causes the redshift distribution to be progressively skewed toward lower redshifts for fainter limiting isophotes.

## 4. Comparison to Observations

The first test of any model that claims to require no evolution is that it match the observed properties of local galaxies. Complete samples of local galaxies, with HI fluxes, colors, scale-lengths, and bulge/disk ratios are non-existent. We are therefore forced to use incomplete samples and to approximate their selection criteria. We have tuned our type-specific luminosity functions as described above to match the locally observed total luminosity function and to be in rough agreement with proportions of morphological types determined by Pence (1976), Tinsley (1980), and Shanks et al. (1984). It is useful to have an independent test that the models reproduce known properties of local galaxies. To this end, in §4.1 and §4.2, we compare our model predictions to local estimates of the HI mass function and galaxy $B - V$ color distribution.

### 4.1. The HI Mass Function

The LSB galaxies that populate our models are for the most part visible on sky-survey plates (that is how many of the known examples have been found), and are especially easy to find if you know where to look (i.e. from an HI position). The impressive constraints on the space density of isolated HI clouds (Fisher & Tully 1981; Briggs 1990) are essentially useless for constraining our model, as the isophotal limits of the searches for optical counterparts have not been quantified. Detections of HI emission in the "off" beams are not uncommon in these surveys (Briggs 1990), but are almost always associated with optically visible galaxies. These galaxies (even if they are LSB) are removed from the samples before the space density of "intergalactic HI clouds" is computed. This HI cloud density therefore does not constrain the LSB galaxy density. An HI mass function derived purely from the off-beam detections without reference to optical counterparts *would* provide strong constraints on our model, but nothing along these lines has been published. The most relevant surveys, therefore, are those that allow some comparison between the frequency of serendipitous detections and the galaxy luminosity function (Kerr & Henning 1987; Weinberg et al. 1991). (The surveys of Schneider et al. (1989) and Hoffman et al. (1989) are in high-density regions, and are

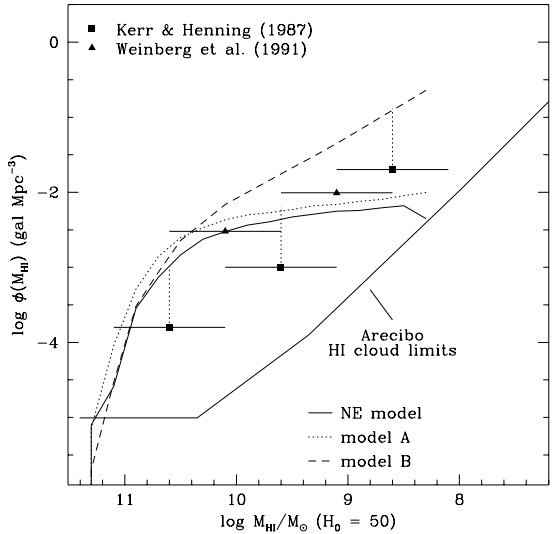

Figure 7: Constraints on the HI mass-function of galaxies. The solid curve shows the HI mass-function for normal galaxies derived from the NE model as described in the text. The dotted and dashed curves show the predictions for models A and B, respectively. Squares show the HI mass function from Kerr & Henning (1987), with dotted lines showing the behavior if the mass-function is normalized to the normal-galaxy HI mass function at $4 \times 10^{10} M_\odot$. The triangles are the Weinberg *et al.* (1991) mass function, normalized to that computed from the canonical optical luminosity function $M_{HI} = 10^{10.1} M_\odot$. The Arecibo HI cloud limits are the limits on the space-density of isolated HI clouds computed by Briggs (1990).

therefore difficult to compare to the expectations from the field-galaxy luminosity function.)

Kerr & Henning (1987) carried out a blind HI search with the 300' Greenbank telescope. They observed 1900 test directions in the galactic plane and 860 directions out of the plane, detecting a total of 28 objects. Briggs (1990) converts their numbers to an HI mass function, which we show in Fig. 7 (converted to proper units for $H_0 = 50$). Comparison of the Kerr & Henning (1987) mass function to the optical luminosity function is highly uncertain, as we do not know the density of galaxies behind the plane of the Milky Way. With our adopted $L_B$ to $M_{HI}$ conversion, objects with masses $M_{HI} > 1 \times 10^9 M_\odot$ appear deficient by a factor of $\sim 7$ compared to the expected HI mass function for normal galaxies. This deficit was also noted by Briggs (1990) and attributed to the difficulty of detecting broad line-width galaxies. As this difficulty is not quantified, we do not know how many large line-width LSB galaxies could be lurking in the night sky. However, the deficit may also be due in part to a difference in mean densities of galaxies behind the plane. If we normalize the Kerr & Henning HI mass function to the field-galaxy Schechter function at $M_{HI} = 10^{10} M_\odot$ (as shown by the dotted lines in Fig. 7), our dwarf dominated model B is still marginally consistent with the data.

Weinberg et al. (1991) conducted another blind HI survey using the VLA. They chose fields both within the Perseus-Pisces supercluster and in a foreground void. Seventeen objects were detected in the supercluster and none in the void. Once again, normalization is a problem and we have chosen simply to normalize the counts in the mass range $10^9 < M_{HI} < 10^{10}$ to the field galaxy $M_{HI}$ mass function computed from our fiducial local luminosity function. This is not strictly proper, as some of the Weinberg et al. fields were centered on bright galaxies. If anything, however, that should lead us to *underestimate* the relative proportion of low $M_{HI}$ galaxies. The Weinberg et al. mass function is shown as triangles in Fig. 7. Weinberg et al. find that their HI mass function is consistent with a flat luminosity function $\alpha = -1$ in the mass range $10^8 < M_{HI} < 10^9$. However, with only 13 galaxies in this mass range, the constraints on the slope are not particularly good. Furthermore, their velocity resolution, 40 km s$^{-1}$, was not optimal for detecting narrow line-width galaxies, and may lead to some incompleteness at the low-mass limits of their mass function.

We conclude that our models are neither supported nor ruled out by existing limits on the HI mass function. The steep HI mass function of model B shows the largest discrepancy with the data, but is still within the uncertainties of the overall normalization of the observed HI mass function and the conversion between blue luminosity and HI mass. As most known LSB field galaxies are rich in HI, a more complete HI survey would provide a stringent test of our proposal.

### 4.2. Local $B - V$ Colors

The choice of SED's for the bulge and disk components in our model was largely dictated by the availability of the Coleman et al. (1980) templates. Our procedure for tuning the luminosity type-specific functions provides no guarantee that the color distribution of galaxies will match that seen in local galaxy samples. Other investigators (Tinsley 1980) have used the $B_J - R_F$ colors distribution of Kirshner et al. (1978) to fix the distribution of morphological types. However, there are difficulties in reproducing the Kirshner et al. (1978) passbands (Bruzual 1981). Instead, we selected all galaxies from the RC3 (de Vaucouleurs et al. 1991) satisfying the following criteria:

$$B_T < 14,\ D_{25} > 120'',\ |b| > 30°,\ \text{and}\ v_{\text{GSR}} < 4000\ \text{km s}^{-1}.$$

To correct for extinction we use the catalogued $A_B$ values and assume $A_B = 4E(B-V)$. We suspect that the RC3 is reasonably complete to this limit. Of those galaxies satisfying the above criteria, 92% have measured $B - V$, sufficient for the color distribution to be representative of the sample as a whole.

We apply the same diameter, magnitude, and velocity selection criteria to our models and compare the resulting color distribution to the RC3 sample in Fig. 8. For comparison, we also show the local B-V distribution that would be inferred from the (Koo et al. 1993) model if galaxies were selected purely by total magnitude.

As our models were not tuned to match the observed color distribution, we regard the agreement for models A and B as satisfactory. Model B shows the best agreement (and also the best agreement with the deep survey data — see below). As our NE model is very similar to other NE models found in the literature, they would presumably show the same surfeit of red galaxies if subjected to the same test. Part of the failure of NE models to match the counts may therefore be due to poor assumptions about the distribution of $z = 0$ SED's, as suggested by Koo et al. (1993). However, the Koo et al. (1993) counter-evolution model appears to skew the distribution too far to the blue.

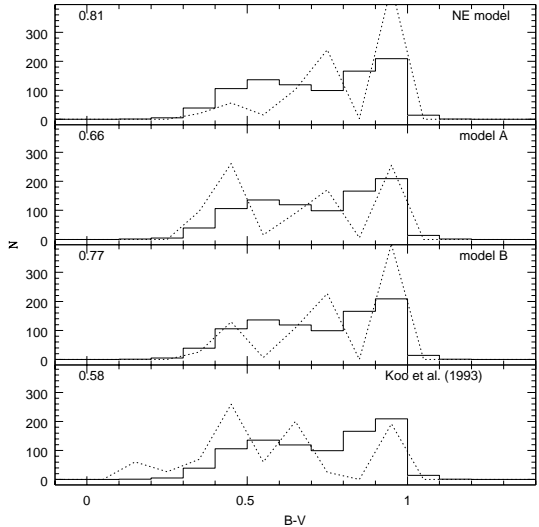

Figure 8: Local $B - V$ color distribution for our models, and for that of Koo et al. (1993). The comparison sample is selected from the RC3 catalog as described in the text. The mean $B - V$ for the different samples are shown in the upper left. The mean for the RC3 sample is $B - V = 0.72$. The models have been normalized to match the number of galaxies in the RC3 sample.

### 4.3. Number Counts

Having shown that our models are at least consistent with low-redshift observations, we now turn to a comparison with estimates of $N(m)$ from deep surveys. Our assumptions about the surveys are found in Table 2. The results are shown in Fig. 9. The NE model with isophotal selection follows quite closely the NE model computed with total magnitude selection. In both cases our NE model agrees closely with that of Yoshii & Takahara (1988), and falls short of the observed counts by a factor of four at $B_J = 24$. Model A is a slight improvement, but still falls short by a factor of three at $B_J = 24$. Model B begins to depart from the observed counts at $B_J = 23$, and is a factor of two short at $B_J = 24$.

The trends are similar in the $I$ band, with model B providing an acceptable fit down to $I_{AB} \sim 23$. In the $K$ band, the agreement between the standard NE model and the observations becomes worse when isophotal selection is included. Once again, model B provides the best fit.

The turnover at faint magnitudes is of course in part due to the adopted ($q_0 = 0.5$) cosmology. For $q_0 = 0.05$ (not shown in the figures), model B agrees with the observed counts to within a factor of two down to $B = 25$ and to within the observational uncertainties down to the limits of the $I$ and $K$ band observations.

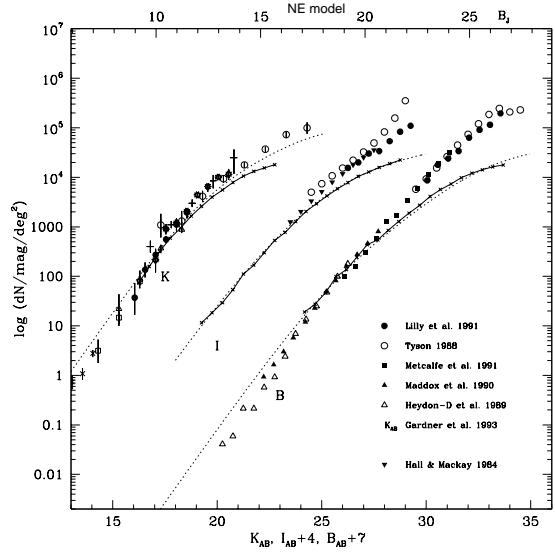

Figure 9: This and the next two figures show number counts in the $B_{AB}$, $I_{AB}$, and $K_{AB}$ bands for our models compared to the data from various sources. Solid lines show the expected $N(m)$ for each of our models. The dashed lines show the standard (non-Monte-Carlo) no-evolution model based on total-magnitude selection as described in the text. This figure shows the NE Model.

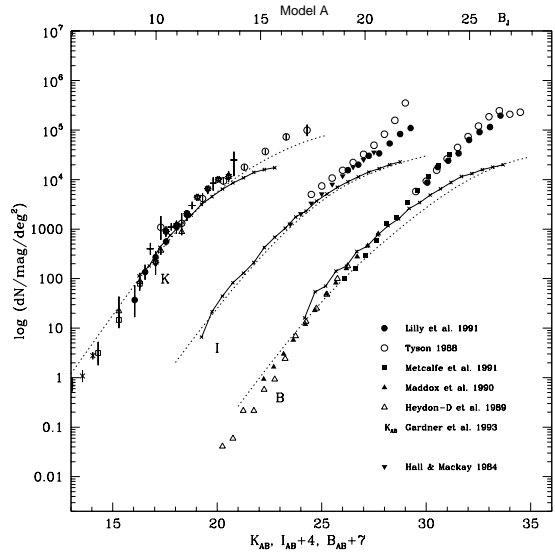

Figure 10: Model A.

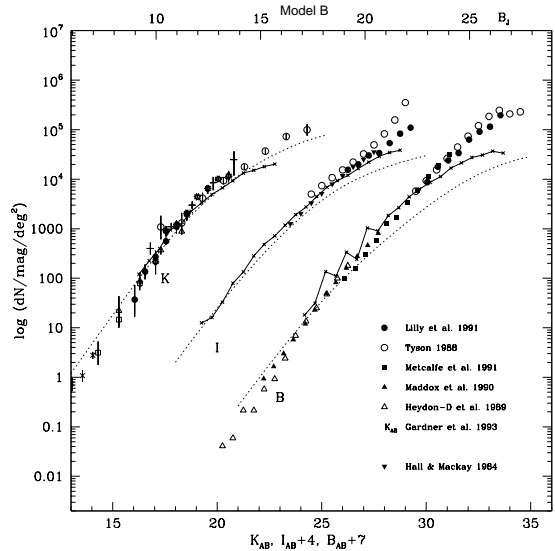

Figure 11: Model B.

Table 4: Median Redshifts

| Survey | Band | Magnitude Range | NE | model A | model B | Observed |
|---|---|---|---|---|---|---|
| Colless | $B_J$ | 21-22.5 | 0.30 | 0.27 | 0.27 | 0.32 |
| Cowie | $B_{AB}$ | 21-24.0 | 0.42 | 0.41 | 0.36 | 0.30 |
| Lilly | $I_{AB}$ | 21-22.5 | 0.52 | 0.51 | 0.54 | 0.38 |

### 4.4. Redshift Distributions

Our model redshift distributions are compared to the observations in Fig. 12. The observed median redshifts are $z = 0.32, 0.30$ and $0.38$ for the Colless, Cowie, and Lilly samples, respectively. The median redshifts predicted by our models are shown in Table 4. The models predict a slightly higher proportion of low redshift galaxies than observed by Colless et al. (1993), and slightly more high redshift galaxies than Cowie sees. The most serious discrepancy is in the $I$ band, where our models predict a median redshift 25% higher than observed. As the models underpredict the counts in the magnitude ranges sampled by the Lilly ($I$-band) and Cowie ($B$-band) surveys, detailed agreement with the observed redshift distributions is not expected. The most important point is that none of the models, when properly normalized, overpredict the numbers of galaxies seen at low-$z$. This was certainly not a foregone conclusion for model B, which is dominated by dwarfs, and illustrates the importance of isophotal selection. The median redshifts are significantly lower for this model when galaxies are selected by total magnitude.

### 4.5. Color Distributions

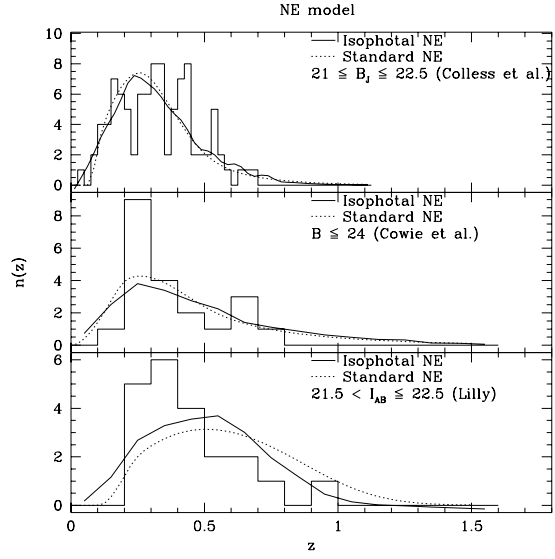

Figure 12: This and the next two figures show redshift distributions in the $B_J$, $B_{AB}$, and $I_{AB}$ bands. The histograms show the data. The solid curve shows the results of the Monte-Carlo models that simulate the selection criteria of the surveys, and the dotted lines show the result of the standard NE model.

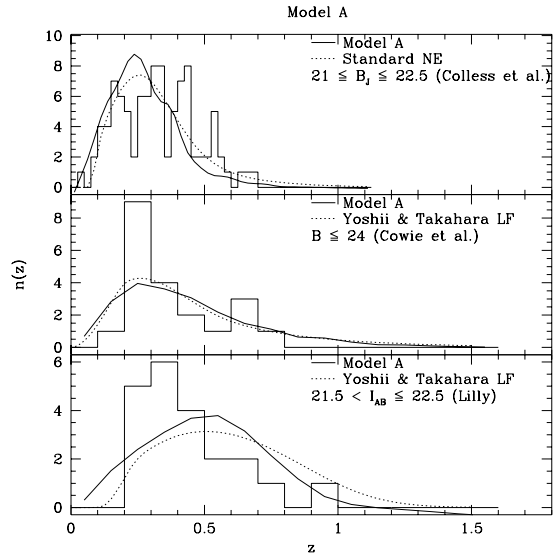

Figure 13: Redshift distributions for model A.

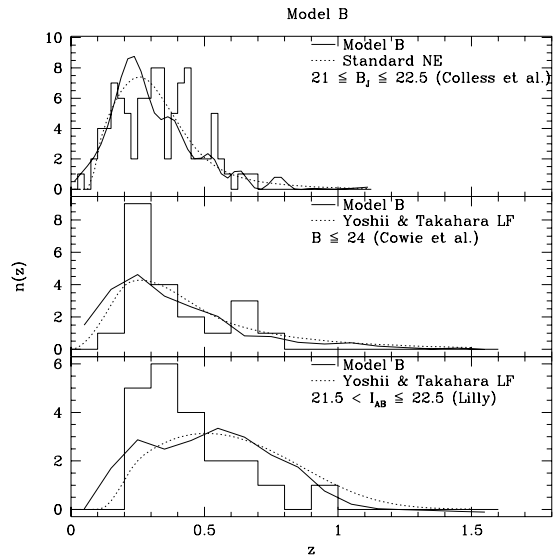

Figure 14: Redshift distributions for model B.

Table 5: $B_{AB} - I_{AB}$ Colors

| Sample | Mean | $\sigma$ |
|---|---|---|
| Lilly (observed) | 1.52 | 0.85 |
| NE model | 1.98 | 1.05 |
| Model A | 1.90 | 1.09 |
| Model B | 1.51 | 0.90 |

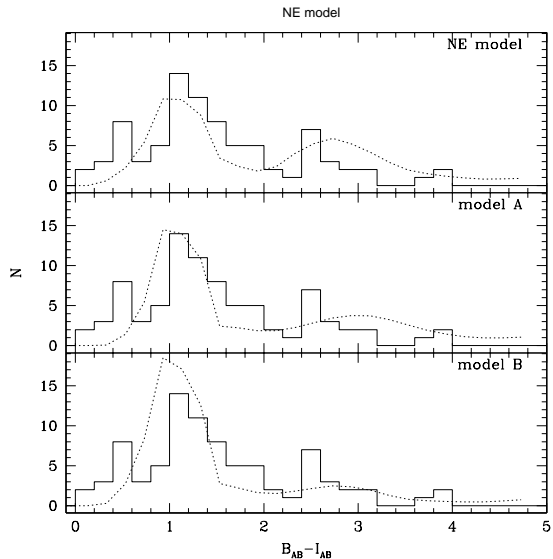

Figure 15: $(B - I)_{AB}$ color distributions for the models limited at $I_{AB} < 25$, compared to the data of Lilly et al. (1991). The models were binned in 0.2 mag intervals, then smoothed with a boxcar filter over 3 bins to simulate photometric errors near the survey limits.

Color distributions can provide an additional constraint on the models. With only three SED's we cannot hope to match the color distribution in detail, however we can at least test whether the mean and dispersion of the predicted colors are close to those observed. In Fig. 15 we compare the model colors to those observed by Lilly, Cowie, & Gardner (1991). Galaxies were required to have $20 < I_{AB} < 25$ and $B_{AB} < 27$ through a 3″ aperture in 1.2″ seeing for both the models and the data. The model color distributions plotted in the figure are are binned in 0.2 mag intervals, and smoothed with a boxcar filter over three bins to simulate photometric errors. The means and standard deviations of the $B_{AB} - I_{AB}$ colors are shown in Table 5.

## 5. Diagnostics for High-Resolution Imaging

One of the major goals of high resolution imaging with the refurbished Hubble Space Telescope (HST) is to reveal the morphologies of high-redshift galaxies. In this section we illustrate the differences between models dominated by LSB galaxies and the standard NE model when the faint galaxies are well resolved. As neither the standard NE model nor the models dominated by LSB galaxies have been tuned (e.g. by adjusting $q_0$) to match the counts at faint magnitudes, detailed agreement is not expected. However, neither changing cosmology nor including mild evolution is likely to change the *relative* distributions of the "normal galaxy" models and LSB galaxy dominated models in the diagrams we present.

As we have not simulated models involving evolution, we cannot compare our predictions for faint-galaxy morphology in detail to those of other models. The only other model to make *quantitative* predictions for faint galaxy morphology is that of Lacey et al. (1993). However, a few very general considerations may serve to illustrate the expected differences between a universe dominated by slowly-evolving LSB galaxies and one dominated by merging galaxies or star-forming dwarfs. In a model dominated by bursting dwarf galaxies, for example, those galaxies that have completed their starburst but have not yet faded from view will be redder than their bursting counterparts. The model therefore predicts that low-surface brightness galaxies will be redder than high-surface brightness galaxies. Our LSB galaxy models predict the opposite trend. Models involving merging and triggered star formation predict that the bluest galaxies will have close neighbors, or will be in the process of merging and therefore have disturbed morphologies. On the other hand, if they are similar to local LSB galaxies, the bluest galaxies in deep surveys should have a dearth of close companions and should display the morphologies of late-type spirals or irregulars.

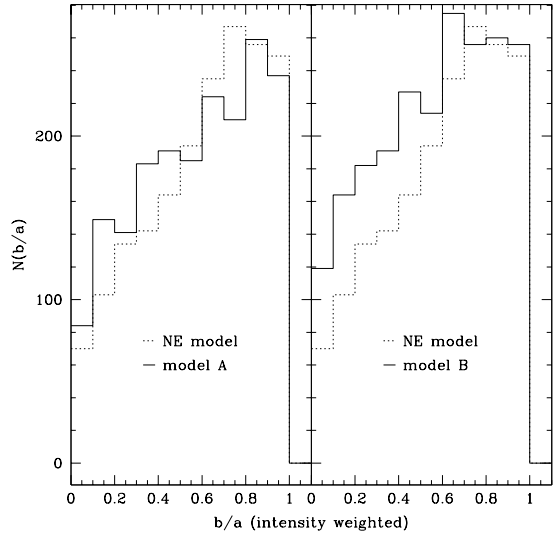

Figure 16: Distribution of intensity-weighted axial ratios ($b/a$). Galaxies for this sample were selected in the $I_{AB}$ band to have diameters $D_{28} > 1''$ and magnitudes $23 < m_{iso} < 25$. The axial ratio is the weighted mean of the bulge and disk axial ratios, using the flux above the limiting isophote as the weighting factor. The instrumental PSF FWHM is assumed to be 0.1″. Smearing by this PSF is included in the modeling the selection of galaxies, but not in the computation of their axial ratios.

As most of the galaxies of interest are near the WFPC-2 detection limit for reasonable exposure times ($< 10$ ks), the task of measuring faint-galaxy morphologies may not be entirely straightforward. Bulge/disk decomposition at low S/N, for example, may not provide stable estimates of the central surface brightnesses of disks. In the remainder of this section we explore several parameters (axial ratios, effective radii, and isophotal magnitudes) that can be measured in WFPC-2 images without detailed profile fitting and that provide a robust distinction between models that are dominated by LSB galaxies and models that are not.

*5.1. Axial Ratio Distribution*

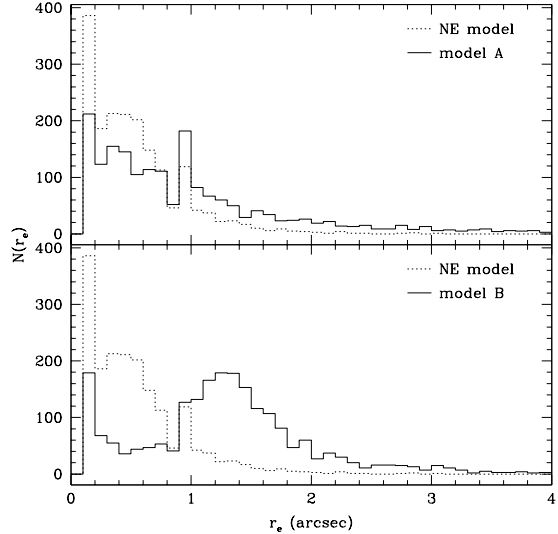

Figure 17: Distribution of half-light radii $r_e$ for the three models. Selection criteria are the same as the previous figure. The computation of $r_e$ includes the effect of smearing by the 0.1″ PSF.

In the LSB galaxy models the counts are dominated by disk galaxies. A simple way of testing whether disk galaxies dominate the counts is to look at the distribution of axial ratios ($b/a$). Figure 16 shows the distribution our three models. Model B is clearly the most dominated by disk galaxies and would be easily distinguished from the no-evolution case in the absence of competing effects. However, there are fairly serious competing effects. Dust in the disks of LSB galaxies could make edge-on galaxies appear fainter than their face-on counterparts. We have included no extinction in our models. Perhaps equally important is the effect of inclination on the selection algorithm. Galaxy detection routines such as FOCAS detect objects by recording the number of connected pixels above a fixed threshold. In the absence of dust, an edge-on LSB galaxy will contain a smaller number of pixels, but each pixel will have a higher flux than the detected pixels in the same galaxy seen face on. Face-on galaxies near the survey limit will tend to break up into many small regions of unconnected pixels, while their edge-on counterparts might not. Finally, in dwarf galaxies the stellar velocity dispersion is a much larger fraction of the rotation velocity than in $L^*$ galaxies, suggesting that the dwarfs might not be as flat as we have assumed. This could be an important effect in the dwarf-dominated model B.

### 5.2. Effective Radii

The distribution of half-light radii provides a much more robust test of models involving LSB galaxies. We have computed $r_e$ from the one-dimensional profiles in our catalogs, including the effects of seeing. Figure 17 shows the predicted $r_e$ distribution for the three models considered here. Lilly et al. (1991) find a distribution of $r_e$ in $V$ images of an $I_{AB}$ selected sample that peaks $\sim 0.5''$ (close to the limit imposed by seeing). The distribution for Model B to roughly the same $I_{AB}$ limit peaks at 1.3″, suggesting that a constant-size relation for galaxies with $L < L^*$ is probably too extreme.

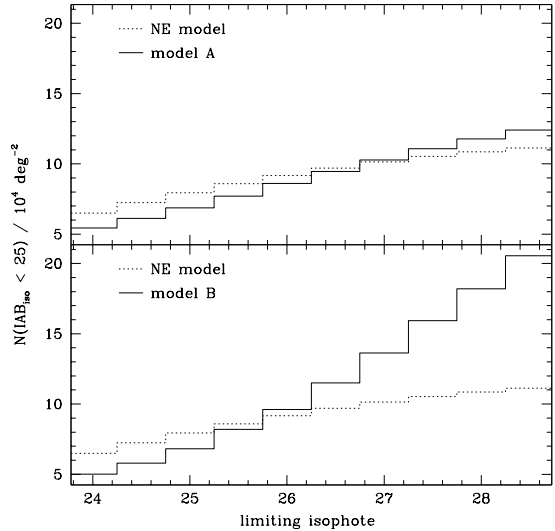

Figure 18: The variation in the number of galaxies detected brighter than an isophotal magnitude of $I_{AB} = 25$ with the limiting isophote. The instrumental PSF is assumed to be 0.1″ FWHM. No diameter limit is imposed on the galaxies.

### 5.3. The Effect of Isophotal Thresholds

Another way to test for LSB galaxies is to examine the effect of imposing different isophotal thresholds on the number counts. In the standard NE model the number of galaxies seen to a fixed isophotal limiting magnitude is not a strong function of the isophotal limit chosen (so long as it is well below the Freeman value). On the other hand, the number of galaxies detected in LSB galaxy models is quite sensitive to the limiting isophote. Figure 18 shows the surface-density of galaxies in our models brighter than an isophotal magnitude $I_{AB} = 25$ as a function of the limiting isophote. The expected number of galaxies rises much more steeply in model B than in the other models.

### 5.4. Simulated Images

The tests described above provide a quantitative means of establishing whether LSB galaxies are a significant contributor faint galaxy counts. The simulated images shown in Figs. 19-21 provide a more qualitative impression of the approximate morphologies expected in HST images. To produce the images, galaxies were selected at random from the simulated catalogs, with the appropriate relative proportions of the different morphological types. Image parameters were were fed to the IRAF [4] "mkobjects" task and the resulting images were convolved with a model WFPC-2 point-spread function. The NE model is shown in Fig. 19, *scaled up by a factor of two in density* to provide a reasonable number of galaxies in the image. Even at HST resolution, many of the galaxies are barely resolved. In contrast, model B (Fig. 21) shows a larger number of galaxies (as it must to match the counts), and a much larger fraction of resolved galaxies. These appear as as fuzzy patches, which, although lacking detailed structures like spiral arms not attempted in the simulations, are clearly distinct from the majority of galaxies seen in the standard NE image. The larger fraction of edge-on systems is also readily apparent. The distinctions between Model A (Fig. 20) and the NE model are more subtle, but still readily apparent to the eye. Ultimately, the most robust test of the models and assessment the potential biases in of various detection algorithms would be to take such simulations and analyze them in the same way as the real observations.

---

[4]IRAF is distributed by Kitt Peak National Observatory, National Optical Astronomy Observatories, operated by the Association of Universities for Research in Astronomy for the

Figure 19: Simulated WFPC-2 (WF3) image of a standard NE model, *with galaxy densities increased by a factor of two*. The area covered is $80 \times 80''$, with a pixel size of $0.1''$. In this and the following images, exposure times were assumed to be $1 \times 10^4$s through the F814W filters. A count rate of $0.039 \mathrm{counts\,s^{-1}\,pixel^{-1}}$ for the sky background, estimated from the WFPC-2 instrument handbook for high ecliptic latitudes. Images were convolved with a model WFPC-2 PSF before adding noise with the IRAF 'mknoise' task. Read noise was taken to be $7e^-\,\mathrm{pixel}^{-1}$, and dark count was assumed to be $0.015\ \mathrm{counts\,s^{-1}\,pixel^{-1}}$. Comparison to figure 21 illustrates the dramatic difference in morphologies expected if LSB galaxies are the dominant contributor to faint galaxy counts.

Figure 20: Simulated WFPC-2 image of model A.

Figure 21: Simulated WFPC-2 image of model B. To give an idea of the depth of the simulated images, two galaxies are marked. $I_{AB}$ magnitudes are 23.8 and 25.1 for galaxies 1 and 2, respectively.

## 6. Conclusions

Our goal in this paper has been to construct a set of alternative models of *non-evolving* galaxy distributions to test whether LSB galaxies could be a significant contributor to faint galaxy counts, and to illustrate the importance of galaxy selection and magnitude measurements schemes in intrepreting faint-galaxy counts.

We have compared two rather extreme models for the surface-brightness distribution of galaxies to a standard non-evolving model. In model A galaxy disks have central surface brightnesses ranging from the Freeman value of $\mu_0(B_J) = 21.6$ to $\mu_0(B_J) = 25$, with constant numbers per unit magnitude independent of luminosity. In contrast, model B incorporates a steep surface-brightness–luminosity relation. Disk galaxies brighter than $L^*$ have $\mu_0(B_J) \sim 21.6$ ($\pm 0.4$ mag), while those fainter than $L^*$ follow a constant size relation (with the same scatter). By tuning the type-specific luminosity functions, we find that we can include a large population of LSB galaxies in either model A or model B without violating the constraints on the local field-galaxy luminosity function (e.g. Loveday et al. 1992). The HI mass function is not in conflict with observations, but could be easily tested with a larger *purely HI* survey, or with more attention to the selection criteria for identifying the optical counterparts of existing HI detections.

The inclusion of LSB galaxies reduces the discrepancy between non-evolving models and faint-galaxy counts for $q_0 = 0.5$. For the parameters of Tyson's (1988) survey, the predicted surface-density of galaxies at $B_J = 24$ for model A is only a factor of 1.2 higher than the NE model. Model B increases the counts by a factor of 2.2, but is still a factor of two shy of the observations at $B_J = 24$. The redshift and color distributions for both models match the observations reasonably well. Agreement between the LSB galaxy models and the faint galaxy data would improve for lower values of $q_0$, or for bluer LSB galaxy colors.

These results indicate that LSB galaxies could *in principle* be an important component of the observed faint-blue-galaxy population. Local examples of LSB galaxies have modest star formation rates, and have probably not drastically changed their luminosities or colors in the last 5 Gyr. For these galaxies, the assumption of no evolution in our models is probably a reasonable approximation. The assumed surface brightness distributions, on the other hand, are completely arbitrary. The distribution of $r_e$ predicted by model B is probably incompatible with ground-based observations, while model A is consistent with the available constraints. The true surface brightness distribution probably lies somewhere between the extremes of our two models.

High-resolution imaging with HST will allow a quantitative estimate of the contribution of LSB galaxies. Sensitive tests for LSB galaxies include measuring the distribution of $r_e$ within a limited magnitude range, and measuring the variation of $N(m)$ as a function of limiting isophote (see §5). Qualitatively, our models differ from models involving merging and/or triggered star formation (Guiderdoni & Rocca-Volmerange 1991; Lacey & Silk 1991; Broadhurst, Ellis, & Glazebrook 1992) in predicting that the bluest galaxies within a fixed magnitude will be the most isolated, and have the lowest surface brightnesses. Models involving the late formation of dwarf galaxies (Babul & Rees 1992) also predict that the bluest galaxies will be the most isolated. However, in these models low-surface-brightness galaxies will be redder on average than high-surface-brightness galaxies, while our models predict the opposite trend.

If LSB galaxies are a significant contributor to the counts of faint galaxies, then the requirements on galaxy evolution models become much less extreme. Standard models of high-surface-brightness galaxy evolution (Bruzual 1983; Arimoto & Yoshii 1987; Rocca-Volmerange & Guiderdoni 1988; Bruzual & Charlot 1993) predict some color and luminosity evolution that must be included in any self-consistent model. We suspect that such modest evolution, coupled with a more realistic surface-brightness distribution and surface-brightness–luminosity relation could restore the agreement between galaxy counts and $q_0 = 0.5$ models, at least to the limits of current redshift surveys.

This work was supported by the Science and Engineering Research Council, and by NASA through grant #HF-1043 awarded by the Space Telescope Science Institute which is operated by the Association of Universities for Research in Astronomy, Inc., for NASA under contract NAS5-26555. We thank Arif Babul for stimulating the Monte-Carlo modeling.